\documentclass[11pt]{article}

\usepackage{amsxtra,amssymb,amsthm,amsmath,latexsym}
\usepackage{graphicx}
\usepackage{epsfig}
\usepackage{afterpage}

\newtheorem{lem}[]{Lemma}

 \newcommand{\lemref}[1]{Lemma~\ref{#1}}

\newcommand{\R}{{\mathbb R}}

\newcommand{\nb}{\nabla}

\newcommand{\dl}{{\delta}}
\newcommand{\bee}{\begin{equation*}}
\newcommand{\eee}{\end{equation*}}
\newcommand{\be}{\begin{equation}}
\newcommand{\ee}{\end{equation}}

\title{Electromagnetic wave scattering by many small particles
and creating materials with a desired permeability}
\author{A.G. Ramm \\
\small Department of Mathematics\\[-0.8ex]
\small Kansas State University, Manhattan, KS 66506-2602, USA\\
\small \texttt{ramm@math.ksu.edu}}

\date{}
\begin{document}
Progress in Electromagnetic Research, M (PIER M, 14, (2010),193-206.
\maketitle

\begin{abstract}
Scattering of electromagnetic (EM) waves by many small imedance particles
(bodies),
embedded in a homogeneous medium, is studied. Physical properties of
the particles are described by their boundary impedances. The
limiting equation is  obtained
for the effective EM field in the limiting medium,
in the limit $a\to 0$, where $a$ is the characteristic size of a
particle and the number $M(a)$ of the particles tends to infinity at
a suitable rate. The proposed theory allows one to create a medium
with a desirable spatially inhomogeneous permeability. The main new
physical
result is the explicit analytical formula for the permeability
$\mu(x)$ of the limiting medium. While the initial medium
has a constant permeability $\mu_0$, the limiting medium, obtained
as a result of embedding many small particles with prescribed boundary
impedances, has a non-homogeneous permeability which is expressed
analytically in terms of the density of the distribution of the
small particles and their boundary impedances. Therefore,
a new physical phenomenon is predicted theoretically, namely, appearance
of a spatially inhomogeneous permeability as a result of embedding of many
small particles whose physical properties are described by their boundary
impedances.
\end{abstract}

{\it PACS}: 02.30.Rz; 02.30.Mv; 41.20.Jb

{\it MSC}: \,\, 35Q60;78A40;  78A45; 78A48;

\noindent\textbf{Key words:} electromagnetic waves; wave scattering
by many small bodies; smart materials.

\section{Introduction}
In this paper we outline a theory of electromagnetic (EM) wave
scattering by many small impedance particles (bodies) embedded in a
homogeneous
medium which is described by the constant permittivity $\epsilon_0>0$,
permeability $\mu_0>0$ and, possibly, constant conductivity $\sigma_0\ge
0$. The small particles are embedded in a finite domain $\Omega$.
The medium, created by  the embedding of the small particles, has new
physical properties.
In particular, it has a spatially inhomogeneous magnetic permeability
$\mu(x)$, which can be controlled by the choice of
the boundary impedances of the embedded small particles and their
distribution density.  This is
a new physical effect, as far as the author knows.
An analytic formula for the permeability of the new medium is derived:
$$\mu(x)=\frac {\mu_0}{\Psi(x)},$$
where
$$\Psi(x)=1+\frac{8\pi i}{3\omega \mu_0}  h(x) N(x).$$
Here $\omega$ is the frequency of the EM field, $\mu_0$ is the
constant permeability parameter of the original medium, $h(x)$ is a
function describing boundary impedances of the small embedded
particles, and $N(x)\geq 0$ is a function describing the
distribution of these particles. We assume that in any subdomain
$\Delta$, the number $\mathcal{N}(\Delta)$ of the embedded particles
$D_m$ is given by the formula:
$$\mathcal{N}(\Delta)=\frac{1}{a^{2-\kappa}}\int_{\Delta}
N(x)dx[1+o(1)],\quad
a\to 0, $$
where $N(x)\geq 0$ is a continuous function, vanishing
outside of the finite domain $\Omega$ in which
small particles (bodies)  $D_m$ are distributed,
$\kappa\in(0,1)$ is a number one can choose at will, and
the boundary impedances of the small particles are defined by the
formula
$$\zeta_m=\frac{h(x_m)}{a^\kappa},\quad x_m\in D_m,$$
where $x_m$ is a point inside $m-$th particle $D_m$, Re $h(x)\geq 0$, and
$h(x)$ is a continuous function vanishing
outside $\Omega.$ The impedance boundary condition on the surface $S_m$
of the $m-$th particle $D_m$ is $E^t=\zeta_m [H^t,N]$, where $E^t$ ($H^t$)
is the tangential component of $E$ ($H$) on $S_m$, and $N$ is the unit
normal to $S_m$, pointing out of $D_m$.

Since one can choose the functions $N(x)$ and $h(x)$, one can create a
desired magnetic permeability in $\Omega$. This is a novel idea, to the
author's knowledge, see also \cite{R581}.

We also derive an analytic
formula for the refraction coefficient of the medium in $\Omega$
created by the embedding of many small particles.
An equation for the EM field in the limiting medium is derived.
This medium is created  when the size $a$ of small particles tends to zero
while the total number $M=M(a)$ of the particles tends to infinity at a
suitable rate.

The refraction coefficient in the limiting medium is spatially
inhomogeneous.

Our theory may be viewed as a "homogenization theory", but it
differs from the usual homogenization theory (see, e.g., \cite{CD},
\cite{MK}, and references therein) in several respects: we do not
assume any periodic structure in the distribution of small bodies,
our operators are non-selfadjoint, the spectrum of these operators
is not discrete, etc. Our ideas, methods, and techniques are quite
different from the usual methods. These ideas are similar to  the
ideas developed in papers \cite{R509, R536}, where scalar wave
scattering by small bodies was studied, and in the papers
\cite{R563},\cite{R598}. However, the scattering of EM waves brought
new technical difficulties which are resolved in this paper. The
difficulties come from the vectorial nature of the boundary
conditions. Our approach is valid for small particles of arbitrary
shapes, but for simplicity we assume that the small bodies are balls
of radius $a$.

We give a new numerical method for solving many-body wave-scattering
problems for small scatterers.

In Section 2 an outline of our theory is given and the basic results are
formulated and explained. In Section 3 the Conclusions are formulated.
In Section 4 short proofs of two lemmas are given. In Section 5, Appendix,
more difficult and lengthy proofs are given,
and a numerical method for solving
many-body wave scattering problem in the case of small scatterers
is presented.

\section{EM wave scattering by many small particles}

We assume that many small bodies $D_m$, $1\leq
m\leq M$, are embedded in a homogeneous medium with constant
parameters $\epsilon_0$, $\mu_0$. Let $k^2=\omega^2\epsilon_0\mu_0$,
where $\omega$ is the frequency. Our arguments remain valid if one
assumes that the medium has a constant conductivity $\sigma_0>0$.
In this case $\epsilon_0$ is replaced by $\epsilon_0+i\frac
{\sigma_0}{\omega}.$
Denote by $[E,H]=E\times H$ the cross
product of two vectors, and by $(E,H)=E\cdot H$ the dot product of two
vectors.

Electromagnetic (EM) wave scattering problem
consists of finding vectors $E$ and $H$ satisfying the Maxwell
equations:
\be\label{e1} \nb \times E=i\omega\mu_0 H,\quad \nb\times
H=-i\omega \epsilon_0 E\quad \text{in } D:=\R^3\setminus
\cup_{m=1}^M D_m, \ee
the impedance boundary conditions:
\be\label{e2} [N,[E,N]]=\zeta_m[H,N]\text{ on
} S_m,\ 1\leq m\leq M, \ee
and the radiation conditions:
\be\label{e3} E=E_0+v_E,\quad H=H_0+v_H,
 \ee
where $\zeta_m$ is the impedance, $N$ is the unit normal to $S_m$
pointing out of $D_m$, $E_0, H_0$ are the incident fields satisfying
equations \eqref{e1} in all of $\R^3$. One often assumes that
the incident wave is a plane wave, i.e.,
$E_0=\mathcal{E}e^{ik\alpha\cdot x}$, $\mathcal{E}$
is a constant vector, $\alpha\in S^2$ is a unit vector, $S^2$ is the
unit sphere in $\R^3$, $\alpha\cdot \mathcal{E}=0$,
$v_E $ and $v_H$ satisfy the
radiation condition: $r(\frac{\partial v}{\partial
r}-ikv)=o(1)$ as $r:=|x|\to \infty$.

By impedance $\zeta_m$ we assume in this paper either a constant,
Re $\zeta_m\geq 0$, or a matrix function $2\times 2$ acting on the
tangential to $S_m$ vector fields, such that \be\label{e4}
\text{Re}(\zeta_mE^t,E^t)\geq 0\quad \forall E^t\in T_m, \ee where
$T_m$ is the set of all tangential to $S_m$ continuous vector fields
such that Div$E^t=0$, where Div is the surface divergence, and $E^t$
is the tangential component of $E$.
Smallness
of $D_m$ means that $ka\ll 1$, where $a=0.5\max_{1\leq m\leq M}
\text{diam} D_m$. By the tangential to $S_m$ component $E^t$ of a vector
field $E$ the following is understood in this paper:
\be\label{e5}
E^t=E-N(E,N)=[N,[E,N]]. \ee This definition differs from the one
used often in the literature, namely, from the definition $E^t=[N,E]$.
Our definition \eqref{e5} corresponds to the geometrical meaning of
the tangential component of $E$ and, therefore, should be used. The
impedance boundary condition is written usually as
$$E^t=\zeta[H^t,N],$$
where the impedance $\zeta$ is a number. If one
uses definition \eqref{e5}, then this condition reduces to
\eqref{e2}, because $[[N,[H,N]],N]=[H,N].$

\begin{lem}\label{lem1}
Problem \eqref{e1}-\eqref{e4} has at most one solution.
\end{lem}
\lemref{lem1} is proved in Section 2.\\
Let us note that problem \eqref{e1}-\eqref{e4} is equivalent to the
problems \eqref{e6}, \eqref{e7}, \eqref{e3}, \eqref{e4}, where
\be\label{e6} \nb\times \nb\times E=k^2E\text{ in } D,\quad
H=\frac{\nb\times E}{i\omega \mu_0}, \ee \be\label{e7}
[N,[E,N]]=\frac{\zeta_m}{i\omega \mu_0}[\nb\times E,N]\text{ on }
S_m,\ 1\leq m\leq M. \ee
Thus, we have reduced our problem to
finding one vector $E(x)$. If $E(x)$ is found, then
$H=\frac{\nb\times E}{i\omega\mu_0}.$

Let us look for $E$ of the
form \be\label{e8} E=E_0+\sum_{m=1}^M\nb \times
\int_{S_m}g(x,t)\sigma_m(t)dt,\quad
g(x,y)=\frac{e^{ik|x-y|}}{4\pi|x-y|}, \ee where $t\in S_m$ and $dt$
is an element of the area of $S_m$, $\sigma_m(t)\in T_m$. This $E$
for any continuous $\sigma_m(t)$ solves equation \eqref{e6} in $D$
because $E_0$ solves \eqref{e6} and
\be\label{e9}\begin{split}
\nb\times\nb\times\nb\times \int_{S_m}g(x,t)\sigma_m(t)dt&=\nb \nb\cdot
\nb\times\int_{S_m}g(x,t)\sigma_m(t)dt\\
&-\nb^2\nb\times
\int_{S_m}g(x,t)\sigma_m(t)dt\\
&=k^2\nb\times \int_{S_m}g(x,t)\sigma_m(t)dt,\quad x\in D.
\end{split}\ee
Here we have used the known identity $div curl E=0,$ valid
for any smooth vector field $E$, and the known formula
\be\label{eG}
-\nb^2 g(x,y)=k^2g(x,y)+\dl(x-y).
\ee
The integral
$\int_{S_m}g(x,t)\sigma_m(t)dt$ satisfies the radiation condition.
Thus, formula \eqref{e8} solves problem \eqref{e6}, \eqref{e7},
\eqref{e3}, \eqref{e4}, if $\sigma_m(t)$ are chosen so that boundary
conditions \eqref{e7} are satisfied.

Define the {\it effective field}
$E_e(x)=E_e^m(x)=E_e^{(m)}(x,a),$ acting on the $m-$th body $D_m$:
\be\label{e10} E_e(x):=E(x)-\nabla\times
\int_{S_m}g(x,t)\sigma_m(t)dt:=E_e^{(m)}(x), \ee where we assume
that $x$ is in a neigborhood of $S_m$, but $E_e(x)$ is defined for
all $x\in \R^3$. Let $x_m\in D_m$ be a point inside $D_m$, and
$d=d(a)$ be the distance between two neighboring small bodies. We assume
that
\be\label{e11} \lim_{a\to 0}\frac{a}{d(a)}=0,\quad \lim_{a\to
0}d(a)=0. \ee We will prove later that $E_e(x,a)$ tends to a limit
$E_e(x)$ as $a\to 0$, and $E_e(x)$ is a twice continuously
differentiable function. To derive an integral equation for
$\sigma_m=\sigma_m(t)$, substitute
$$E=E_e+\nb\times\int_{S_m}g(x,t)\sigma_m(t)dt$$ into impedance boundary
condition \eqref{e7}, use the known formula (see, e.g., \cite{M}):
\be\label{e12} [N,\nb\times
\int_{S_m}g(x,t)\sigma_m(t)dt]_{\mp}=\int_{S_m}
[N_s,[\nb_xg(x,t)|_{x=s},\sigma_m(t)]]dt\pm \frac{\sigma_m(s)}{2}, \ee
where the $\pm$ signs
denote the limiting values of the left-hand side of \eqref{e12} as $x\to
s$ from
$D$ $(D_m)$, and get the following equation:
\be\label{e13} \sigma_m(t)=A_m\sigma_m+f_m,\quad
1\leq m\leq M. \ee
Here $A_m$ is a linear Fredholm-type integral
operator, defined  by formula $A_m=-2[N_s,B_m\sigma_m]$, 
where the operator $B_m$ is defined  by formula  \eqref{e19},
and $f_m$ is a continuously differentiable vector function, 
defined by formula \eqref{e14}.

Let us find formulas for  $A_m$ and $f_m$. Equation \eqref{e13} is 
derived in  Appendix and there the formulas for $f_m$ and $A_m$ are 
obtained.

One has:
\be\label{e14}
f_m=2[f_e(s), N_s],\quad
f_e(s):=[N_s,[E_e(s),N_s]]-\frac{\zeta_m}{i\omega \mu_0}[\nb\times
E_e,N_s]. \ee
Boundary condition \eqref{e7} and formula \eqref{e12} yield
\be\label{e15}\begin{split}
&f_e(s)+\frac{1}{2}[\sigma_m(s),N_s]+
[\int_{S_m}[N_s,[\nb_sg(s,t),\sigma_m(t)]]dt,N_s]\\
&-\frac{\zeta_m}{i\omega
\mu_0}[\nb\times\nb\times\int_{S_m}g(x,t)\sigma_m(t)dt,N_s]|_{x\to
s}=0.\end{split}\ee

Using the known formula $\nb\times\nb\times =grad div  -\nb^2$, the
relation
\be\label{e16}\begin{split}
\nb_x\nb_x \cdot \int_{S_m}g(x,t)\sigma_m(t)dt&=\nb_x\int_{S_m}(-\nb_t
g(x,t),\sigma_m(t))dt\\
&=\nb_x\int_{S_m}g(x,t)\text{Div}\sigma_m(t)dt=0, \end{split}\ee
where Div is the surface divergence, and the formula
\be\label{e17}
-\nb_x^2\int_{S_m}g(x,t)\sigma_m(t)dt=k^2\int_{S_m}g(x,t)\sigma_m(t)dt,\quad
x\in D, \ee
where  equation  \eqref{eG}  was
used, one gets from \eqref{e15} the following equation
\be\label{e18} -[N_s,\sigma_m(s)]+2f_e(s)+2B\sigma_m=0. \ee
Here
\be\label{e19} B\sigma_m:=[\int_{S_m}[N_s,[\nb_s
g(s,t),\sigma_m(t)]]dt,N_s]+\zeta_mi\omega
\epsilon_0[\int_{S_m}g(s,t)\sigma_m(t)dt,N_s]. \ee
Take cross
product of $N_s$ with the left-hand side of \eqref{e18} and use the
formulas $N_s\cdot \sigma_m(s)=0$, $f_m:=f_m(s):=2[f_e(s), N_s]$, and
\be\label{e20}
[N_s,[N_s,\sigma_m(s)]]=-\sigma_m(s), \ee
to get from \eqref{e18}
equation \eqref{e13}:
\be\label{e21}
\sigma_m(s)=2[f_e(s), N_s]-2[N_s,B\sigma_m]:=A_m\sigma_m+f_m, \ee
where
$$A_m\sigma_m=-2[N_s,B\sigma_m].$$
The operator $A_m$ is linear and compact
in the space $C(S_m)$, so that equation \eqref{e21} is of Fredholm
type. Therefore, equation \eqref{e21} is solvable for any $f_m\in
T_m$ if the homogeneous version of \eqref{e21} has only the trivial
solution $\sigma_m=0$. In this case the solution $\sigma_m$
to equation \eqref{e21} is of the
order of the right-hand side $f_m$, that is, $O(a^{-\kappa})$ as $a\to
0$, see formula \eqref{e14}. Moreover, it follows from equation
\eqref{e21} that the main term of the asymptotics of $\sigma_m$
as $a\to 0$ does not depend on $s\in S_m$.

\begin{lem}\label{lem2}
Assume that $\sigma_m\in T_m, $ $\sigma_m\in C(S_m)$, and
$\sigma_m(s)=A_m\sigma_m$. Then $\sigma_m=0$.
\end{lem}
\lemref{lem2} is proved in Section 2.

Let us assume that in any subdomain $\Delta$, the number
$\mathcal{N}(\Delta)$ of the embedded bodies $D_m$ is given by the
formula:
\be\label{e22}
\mathcal{N}(\Delta)=\frac{1}{a^{2-\kappa}}\int_{\Delta}N(x)dx[1+o(1)],\quad
a\to 0, \ee
where $N(x)\geq 0$ is a continuous function, vanishing
outside of a finite domain $\Omega$ in which
small bodies  $D_m$ are distributed,
$\kappa\in(0,1)$ is a number one can choose at will. We also assume that
\be\label{e23}
\zeta_m=\frac{h(x_m)}{a^\kappa},\quad x_m\in D_m, \ee where
Re $h(x)\geq 0$, and $h(x)$ is a continuous function vanishing outside
$\Omega.$

Let us write \eqref{e8} as
\be\label{e24}
E(x)=E_0(x)+\sum_{m=1}^M[\nb_x g(x,x_m),Q_m]+\sum_{m=1}^M
\nb\times\int_{S_m}(g(x,t)-g(x,x_m))\sigma_m(t)dt, \ee
where
\be\label{e25} Q_m:=\int_{S_m}\sigma_m(t)dt.
\ee
The central physical idea of the theory, developed in this paper,
is simple: one can neglect the second sum in \eqref{e24}
compared with the first sum.

Since $\sigma_m=O(a^{-\kappa})$, one has $Q_m=O(a^{2-\kappa})$.
We want to prove
that the second sum in \eqref{e24} is negligible compared with the
first one. This proof is based on several estimates.

We assume in these estimates that $a\to 0$, $\frac{a}{d}\to 0$,
and $|x-x_m|\ge d$.
Under these assumptions one has
\be\label{e26} j_1:=|[\nb_x g(x,x_m),Q_m]|\leq
O\left(\max\left\{\frac{1}{d^2},\frac{k}{d}\right\}\right)O(a^{2-\kappa}),
\ee
\be\label{e27}
j_2:=|\nb\times\int_{S_m}(g(x,t)-g(x,x_m))\sigma_m(t)dt|\leq a
O\left(\max\left\{\frac{1}{d^3},\frac{k^2}{d}\right\}\right)O(a^{2-\kappa}),
\ee
and
\be\label{e28} \left| \frac{j_2}{j_1}\right|=O\left( \max
\left\{ \frac{a}{d},ka\right\}\right)\to 0,\qquad \frac a d=o(1),  \qquad
a\to 0.  \ee
These estimates show that one may neglect the second sum in \eqref{e24},
and write
\be\label{e29} E_e(x)=E_0(x)+\sum_{m=1}^M[\nb_xg(x,x_m),Q_m] \ee
with an error that tends to zero under our assumptions as $a\to 0$,
and when $|x-x_j|\sim a$ then the term with $m=j$ in the sum
\eqref{e29} should be dropped according to the definition of
the effective field. We will show that the limit of the effective field,
as $a\to 0$ does exist and solves equation \eqref{e37}, see below.

{\it Let us estimate $Q_m$
asymptotically, as $a\to 0$.}

Integrate equation \eqref{e21} over
$S_m$ to get
\be\label{e30}
Q_m=2\int_{S_m}[f_e(s), N_s]ds-2\int_{S_m}[N_s,B\sigma_m]ds .\ee
We will show in the Appendix  that the second term in the right-hand side
of the above equation is equal to $-Q_m$ plus terms negligible compared
with $|Q_m|$ as $a\to 0$. Thus,
\be\label{eQ_m}Q_m=\int_{S_m}[f_e(s), N_s]ds, \qquad a\to 0.\ee
{\it Let us estimate the integral in the right-hand side of \eqref{eQ_m}.
}

It follows from equation \eqref{e14} that
\be\label{e31}
[N_s,f_e]=[N_s,E_e]-\frac{\zeta_m}{i\omega \mu_0}[N_s,[\nb \times
E_e,N_s]]. \ee
If $E_e$ tends to a finite limit as $a\to 0$, then
formula \eqref{e31} implies that \be\label{e32}
[N_s,f_e]=O(\zeta_m)=O\left(\frac{1}{a^\kappa}\right),\quad a\to 0.
\ee
By \lemref{lem2}, the operator $(I-A_m)^{-1}$ is bounded, so
$\sigma_m=O\left(\frac{1}{a^\kappa}\right)$, and \be\label{e33}
Q_m=O\left(a^{2-\kappa}\right),\quad a\to 0, \ee
because the integration over $S_m$ adds factor $O(a^2)$. 
It will follow from our arguments that $Q_m$ does not vanish
at almost all points, see formulas  \eqref{e35}- \eqref{e36}. 

As $a\to 0$,
the sum \eqref{e29} converges to the integral
\be\label{e34}
E(x)=E_0(x)+\nb\times\int_{\Omega}g(x,y)N(y)Q(y)dy, \ee
where $Q(y)$ is the function uniquely defined by the formula
\be\label{e35} Q_m=Q(x_m)a^{2-\kappa}. \ee
The function $Q(y)$ is defined uniquely, because,as $a\to 0$ the set
of points $\{x_m\}_{m=1}^M$ becomes dense in $\Omega$.
The physical meaning of vector $E(x)$ in equation \eqref{e34} is
clear: this vector is the limit of the effective field $E_e(x)$ as $a\to
0$, and $N(x)$ is the function from equation \eqref{e22}.

The function $Q(y)$ can be expressed in terms of $E$:
\be\label{e36}
Q(y)=-\frac{8\pi i}{3\omega \mu_0} h(y) (\nb\times
E)(y). \ee
This important formula is derived in Appendix.

The factor $\frac{8\pi}{3}$ appears if $D_m$ are balls.
Otherwise a tensorial factor $c_{m}$, depending on the shape of $S_m$,
should be used in place of $\frac{8\pi}{3}$.

From equations \eqref{e34} and  \eqref{e36} one obtains
\be\label{e37} E(x)=E_0(x)-\frac{8\pi i}{3\omega \mu_0}
 \nb\times \int_{\Omega}g(x,y)h(y)N(y)\nb\times E(y)dy. \ee

{\it Let us derive physical conclusions from equation
\eqref{e37}.}

Applying  the operator  $\nb\times\nb\times$ to \eqref{e37}
yields
\be\label{e38}\begin{split} \nb\times\nb\times E&=k^2E_0(x)\\
&-\frac{8\pi i}{3\omega \mu_0}  \nb\times (\text{grad
div}-\nb^2)\int_{\Omega}g(x,y)h(y)N(y)\nb\times E(y) dy\\
&=k^2E_0-k^2 \frac{8\pi i}{3\omega \mu_0}
\nb\times\int_{\Omega}g(x,y)h(y)N(y) \nb\times E(y)dy\\
&-\frac{8\pi i}{3\omega \mu_0}h(x)N(x) \nb \times \nb\times 
E(x)\\
&=k^2E(x)-\frac{8\pi i}{3\omega \mu_0}  h(x)N(x) \nb\times\nb\times
E\\
&-\frac{8\pi i}{3\omega \mu_0} [\nb (h(x)N(x)),\nb\times E(x)].
\end{split}\ee
Here we have used  formula $\nb\times \text{grad}=0$, equation
\eqref{eG}, and assumed for simplicity that $h(x)$ is a scalar
function. It follows from \eqref{e38} that \be\label{e39}
\nb\times\nb\times E=K^2(x)E-\frac{\frac{8\pi i}{3\omega
\mu_0}}{1+\frac{8\pi i}{3\omega \mu_0} h(x)N(x)}[\nb
(h(x)N(x)),\nb\times E(x)], \ee where \be\label{e40}
K^2(x)=\frac{k^2}{1+\frac{8\pi i}{3\omega\mu_0} h(x)N(x)},\quad
k^2=\omega^2 \epsilon_0 \mu_0. \ee If
$$\nb \times E=i\omega \mu (x) H, \qquad \nb \times H=-i\omega
\epsilon (x) E,$$ then
\be\label{e41}\nb \times \nb \times E=\omega^2 \epsilon (x) \mu (x) E
+[\frac {\nb \mu (x)}{\mu(x)}, \nb \times E].
\ee
Comparing this equation with \eqref{e39}, one can identify the last
term in \eqref{e39} as coming from a {\it variable permeability} $\mu
(x)$.
This $\mu (x)$ appears in the limiting medium due to the boundary currents
on the surfaces $S_m$, $1\le m \le M$. These currents appear because of
the impedance boundary conditions \eqref{e7}.

{\it Let us identify the permeability $\mu (x)$.}

  Denote
$$\Psi (x):=1+\frac{8\pi i}{3 \omega \mu_0} h(x) N(x).$$ Let
$\epsilon (x)=\epsilon_0$, $\epsilon_0= const,$ and define
$$\mu(x):=\frac {\mu_0}{\Psi (x)}.$$ Then
$K^2=\omega^2 \epsilon_0 \mu (x)$, and $$\frac {\nb \mu
(x)}{\mu(x)}=-\frac {\nb \Psi(x)}{\Psi (x)}.$$ Consequently, formula
\eqref{e39} has a clear physical meaning: the electromagnetic
properties of the limiting medium are described by the variable
permeability: \be\label{e42} \mu(x)=\frac{\mu_0}{\Psi
(x)}=\frac{\mu_0}{1+\frac{8\pi i}{3 \omega \mu_0} h(x) N(x)}. \ee

\section{Conclusions}

{\it The limiting medium
is described by the new refraction coefficient $K^2(x)$ (see \eqref{e40})
and the new term
in the equation \eqref{e39}. This term is due to the spatially
inhomogeneous permeability
$\mu(x)=\frac {\mu_0}{\Psi (x)}$ generated in the limiting medium
by the boundary impedances.
The field $E(x)$ in the limiting medium
( and in equation \eqref{e39}) solves equation \eqref{e37}.

Therefore, we predict theoretically the new physical phenomenon:
by embedding many small particles with suitable boundary impedances
into a given homogeneous medium, one can create a medium with a desired
spatially inhomogeneous permeability \eqref{e42}.

One can create material with a desired permeability $\mu(x)$
by embedding small particles with suitably chosen boundary impedances.
Indeed, by formula \eqref{e42} one can choose a complex-valued, in
general, function $h(x)$, and a non-negative function  $N(x)\ge 0$,
describing the density distribution of the small particles,
so that the right-hand side of formula \eqref{e42} will yield a desired
function $\mu(x)$.  }

\section{Proofs of Lemmas}
{\it Proof of \lemref{lem1}.}\\
\noindent From  equations \eqref{e1} one derives (the bar stands for
complex conjugate):
$$\int_{D_{R}}(\overline{H}\cdot \nb\times E-E\cdot \nb
\times \overline{H})dx=\int_{D_R}(i\omega \mu_0 |H|^2-i\omega
\epsilon_0|E|^2)dx,$$ where $D_R:=D\cap B_R$, and $R>0$ is so large that
$D_m\subset B_R:=\{x\ : \ |x|\leq R\}$ for all $m$.
Recall that $\nb\cdot [E, \overline{H}]=\overline{H}\cdot\nb\times E
-E\cdot\nb\times \overline{H}$.
Applying the divergence theorem, using the radiation condition on the
sphere $S_R=\partial B_R$, and taking real part, one gets
$$0=\sum_{m=1}^M\text{Re}\int_{S_m}[E,\overline{H}]\cdot N
ds=\sum_{m=1}^M\text{Re}\int_{S_m}\overline{\zeta_m}^{-1}\overline{E}^-_t
\cdot E^-_tds,$$
where $E^-_t$ is the limiting value of $E^t$ on $S_m$ from $D$,
$E^t=\zeta_m[H,N]$. This relation and assumption \eqref{e4} imply
$E^-_t=0$ on $S_m$ for all $m$. Thus, $E=H=0$ in $D$.\\
\lemref{lem1} is proved. \hfill $\Box$

{\it Proof of \lemref{lem2}.}

If $\sigma_m=A_m\sigma_m$, then the functions
$$H=\frac{\nb\times E}{i\omega \mu_0}, \qquad E(x)=\nb\times
\int_{S_m}g(x,t)\sigma(t)dt$$
solve equation \eqref{e1} in $D$, $E$ and $H$ satisfy the radiation
condition, and , condition \eqref{e2}. Thus, $E=H=0$ in $D$.

Consequently,
\bee\begin{split} 0&=\nb\times\nb
\times\int_{S_m}g(x,t)\sigma_m(t)dt=(\text{grad
div}-\nb^2)\int_{S_m}g(x,t)\sigma_m(t)dt\\
&=k^2\int_{S_m}g(x,t)\sigma_m(t)dt,\quad x\in D.\end{split}\eee This
implies $\sigma_m(s)=0$.\\
\lemref{lem2} is proved. \hfill $\Box$

\section{Appendix}

In Section 5.1 equation \eqref{e37} is derived. In Section 5.2
a linear algebraic system (LAS) is derived for finding vectors $Q_m$
in equation \eqref{e35}. 
If one substitutes
formula \eqref{e36} into \eqref{e34}, then one obtains equation
\eqref{e37}.

5.1. {\bf Derivation of the basic equation \eqref{e37}}

Boundary condition \eqref{e7} yields \bee\begin{split}
0&=[N[E_e,N]]-\frac{\zeta_m}{i\omega \mu_0}[\nb\times
E_e,N]+[N,[\nb\times\int_{S_m}g(s,t)\sigma_m(t)dt,N]]\\
&-\frac{\zeta_m}{i\omega \mu_0}[\nb\times \nb\times
\int_{S_m}g(x,s)\sigma_m(t)dt,N].\end{split}\eee
Let us denote
$$f_e:=[N,[E_e,N]]-\frac{\zeta_m}{i\omega\mu_0}[\nb\times E_e,N].$$
One has $\nb \times \nb \times=curl curl=grad div - \Delta$,
and
\bee
\nb_x\cdot \int_{S_m}g(x,t)\sigma_m(t)dt=-\int_{S_m}\left(\nabla_t
g(x,t),\sigma_m(t) \right) dt=\int_{S_m} g(x,t)\nb_t\cdot
\sigma_m(t)dt=0,
\eee
and
\bee
-\nb_x^2\int_{S_m}g(x,t)\sigma_m(t)dt=k^2\int_{S_m}g(x,t)\sigma_m(t)dt,
\eee because $-\nb_x^2 g(x,t)=k^2g(x,t)$, $x\neq t$, see \eqref{eG}.
Thus, using
\eqref{e12}, one gets: \bee\begin{split}
0&=f_e+[\int_{S_m}[N_s,[\nb_s
g(s,t),\sigma_m(t)]]dt,N_s]+\frac{1}{2}[\sigma_m(s),N_s]\\
&+\frac{\zeta_m k^2}{i\omega
\mu_0}[N_s,\int_{S_m}g(s,t)\sigma_m(t)dt]. \end{split}\eee
Cross multiply this by $N_s$ from the left and use the relation $N_s\cdot
\sigma_m(s)=0$, to obtain
\bee\begin{split} 0&=[N_s, f_e]+[N_s,[\int_{S_m}[N_s,[\nb_s
g(s,t),\sigma_m(t)]]dt,N_s]]+\frac{1}{2}\sigma_m(s)\\
&-\zeta_m i\omega \epsilon_0[N_s,[N_s,\int_{S_m}g(s,t)\sigma_m(t)dt]].
\end{split}\eee
Note that
\bee\begin{split} [N_s,[\int_{S_m}[N_s,[\nb_s
g(s,t),\sigma_m(t)]]dt,N_s]]&=\int_{S_m}[N_s,[\nb_s
g(s,t),\sigma_m(t)]]dt\\
&-[N_s,N_s]\int_{S_m}\left(N_s,[\nb_s
g(s,t), \sigma_m(t)]dt\right)\\
&=\int_{S_m}[N_s,[\nb_s g(s,t),\sigma_m(t)]]dt.
\end{split},
\eee
Consequently,
\bee\begin{split} \sigma_m(t)&=2[f_e(s),N_s]+2\zeta_m
i\omega \epsilon_0 [N_s,[N_s,\int_{S_m}g(s,t)\sigma_m(t)dt]]\\
&-2\int_{S_m}[N_s,[\nb_sg(s,t),\sigma_m(t)]]dt:=A\sigma_m+f_m,
\end{split}\eee
which is equation \eqref{e13}, and $f_m:=2[f_e(s),N_s],$
which is  equation \eqref{e14}. 

Denote
$$Q_m=\int_{S_m}\sigma_m(s)ds.$$
One has
\bee\begin{split}
\int_{S_m}[[N_s,[E_e(s),N_s]],N_s]ds&=
\int_{S_m}[E_e(s),N_s]ds=
-\int_{D_m}\nb_x\times E_e dx.
\end{split}\eee
The term  $\int_{D_m}\nb_x\times E_e dx=O(a^3)$ is negligible
compared with the terms of order $O(a^{2})$.
Let us estimate the terms of the order  $O(a^{2})$. One has
\bee\begin{split} & \int_{S_m}[[\nb\times E_e,N_s],N_s]ds
=-\left( \int_{S_m}\nb\times E_e ds-\int_{S_m}N_s(\nb\times
E_e,N_s)ds\right)\\
&=-\int_{S_m}\nb\times E_e ds+\frac{4\pi a^2}{3}\nb\times E_e(x_m)\\
&=-\frac{8\pi a^2}{3}\nb\times E_e (x_m), \qquad a\to 0.
\end{split}\eee
Here we have used the formulas
$$\int_{S_m}\nb\times E_e ds\sim 4\pi a^2 \nb\times E_e (x_m),\quad a\to
0,$$
and
$$\int_{S}N_iN_jds=\frac{4\pi a^2}{3}\delta_{ij},$$
where $S$ is a sphere of radius $a$, $\{N_i\}_{i=1}^3$  are Cartesian
components of the outer unit normal to the sphere $S$, and
$\delta_{ij}=0$ if $i\neq j$, $\delta_{ii}=1$.

Thus,
\be\label{eA.Q0}
 0.5\int_{S_m}f_m(s)ds\sim -\frac{8\pi i}{3\omega \mu_0}\zeta_m a^2
\nb\times E_e(x_m)=O(a^{2-\kappa}), \quad  a\to 0 \ee provided that
$$\zeta_m=\frac{h(x_m)}{a^\kappa},\qquad 0<\kappa< 1.$$
Let us now show that the term $\int_{S_m}A\sigma_m ds$ contributes
the term $-Q_m$, so
\be\label{eA.Q}
Q_m= 0.5\int_{S_m}f_m(s)ds \left(1+o(1)\right), \qquad   a\to 0.
\ee
One has
\bee\begin{split} &-2\int_{S_m}ds\int_{S_m}[N_s,[\nb_s
g(s,t),\sigma_m(t)]]dt\\
&=-2\int_{S_m}ds \int_{S_m}dt\left(\nb_s
g(s,t)(N_s,\sigma_m(t))-\sigma_m(t)\frac{\partial g(s,t)}{\partial
N_s} \right)
dt\\
&=-2\int_{S_m}ds \int_{S_m}dt \nb_s
g(s,t)(N_s,\sigma_m(t))+\int_{S_m}\sigma_m(t)dt 2\int_{S_m}ds
\frac{\partial g(s,t)}{\partial N_s}.
\end{split}\eee
Since
$$2\int_{S_m}ds \frac{\partial g(s,t)}{\partial
N_s}=-2\int_{D_m}dxk^2g(x,t)-1,$$
one gets
$$I:=\int_{S_m}dt
\sigma_m(t) 2\int_{S_m}ds\frac{\partial g(s,t)}{\partial
N_s}=-\int_{S_m}\sigma_m(t)dt-2k^2\int_{S_m}dt
\sigma_m(t)\int_{D_m}dx g(x,t).$$
Therefore
$$I:=-Q_m+I_1,$$
where the term $I_1$ is negligible compared with $Q_m$, because
$$\int_{D_m}dx g(x,t)=O(a^2),\qquad a\to 0.$$ Consequently, $I_1$
 is negligible compared with $I$.

If
$\int_{S_m}|\sigma_m(t)|dt<\infty$ and $Q_m=\int_{S_m}\sigma_m(t)dt\neq
0$, then
$$|\int_{S_m}\sigma_m(t)dt|\gg |\int_{S_m}dt
\sigma_m(t)\int_{D_m}dx g(x,t)|,$$
because $|\int_{D_m}dx g(x,t)|=O(a^2)$ if $x\in D_m$.
If $ka\ll 1$,
then the effective field $E_e$ and $\nb \times E_e$ change negligibly at
the distances of order $a$,
and  $Q_m$ is proportional to $a^{2-\kappa} \nb \times
E_e(x_m)$ on the surface $S_m$, and therefore $Q_m\neq 0$ 
at all points at which $\nb \times E_e(x_m)$ does not vanish,
i.e., at almost all points in $\Omega$.

One has
$$ \big{|}-2\int_{S_m}ds\int_{S_m}dt
\nb_sg(s,t)(N_s,\sigma_m(t))\big{|}\ll
\big{|}\int_{S_m}\sigma_m(t)dt\big{|}=|Q_m|,$$
because $|(N_s,\sigma_m(t))|=O(|s-t|)$ as $|s-t|\to 0$.

Therefore,
\be\label{eA0}
Q_m=0.5\int_{S_m}f_m(t)dt\sim -\frac{8\pi i}{3\omega \mu_0}\zeta_m a^2
 \nb\times E_e(x_m) ,\quad a\to 0.
\ee
This yields the following formula (cf \eqref{e29}):
\be\label{eA1} E_e(x)=E_0(x)- \frac{8\pi i}{3 \omega\mu_0}\sum_{m=1}^M
\zeta_m a^2[\nabla g(x,x_m), \nb\times E_e(x_m)], \qquad a\to 0,
\ee
which can be written as
\be\label{eA2}
E_e(x)=E_0(x)- \frac{8\pi i}{3 \omega\mu_0}
a^{2-\kappa}\sum_{m=1}^Mh(x_m)
\left[ \nb_x g(x,x_m), \nb\times E_e (x_m)\right],
\ee
and when $|x-x_j|\sim a$ then the term with $m=j$ in the sum
\eqref{eA2} should be dropped according to the definition of the 
effective field $E_e(x)$. 

Passing to the limit $a\to 0$ in equation \eqref{eA2},
and denoting by $E(x)$ the limit of the field $E_e(x)$ as
$a\to 0$,  one obtains
\be\label{eA3}
E=E_0(x)- \frac{8\pi i}{3 \omega\mu_0}\int_{\Omega} [\nb_x
g(x,y),\nb \times E_e(y)] h(y)N(y)dy,
\ee
where $h(x)$ is the function in the formula $\zeta_m=\frac
{h(x_m)}{a^{\kappa}}$, and $N(x)$ is the function in the definition
of $\mathcal{N}(\Delta)$.

The above passage to the limit is explained in 
Appendix, see also \cite{R598}, p. 206.
This passage uses the convergence of the collocation method for
solving equation \eqref{e37}, see \cite{R563}.
Equation \eqref{eA3} for the limiting field $E(x)$ is equivalent to 
equation \eqref{e37}.

5.2. {\bf Derivation of a linear algebraic system for $\mathcal{P}_m$}

In this Section a numerical method is developed for solving
many-body wave scattering problem when the scatterers are small in
comparison with the wavelength. The method consists of a derivation
of a linear algebraic system for finding vectors
$$\mathcal{P}_m:=(\nb\times E)(x_m),\qquad 1\le m \le M.$$ 
If
$\mathcal{P}_m$ are found, then by formulas \eqref{e36} and
\eqref{e35} one finds \be\label{eA4}Q_m=- \frac{8\pi i}{3
\omega\mu_0}a^{2-\kappa} h(x_m) \mathcal{P}_m, \ee and, by formula
\eqref{e29}, the field $E(x)$.

Let us derive linear algebraic system for finding
$\mathcal{P}_m$.

Apply $\nb \times$ to equation \eqref{eA2}, let $x=x_j$,
$1\le j \le M,$ and replace $\sum_{m=1}^M$ by the sum $\sum_{m\neq
j, m=1}^M$.

Then one obtains
\be\label{eA5}\mathcal{P}_j=\mathcal{P}_{0j}- \frac{8\pi i}{3
\omega\mu_0} a^{2-\kappa} \sum_{m\neq
 j, m=1}^M \{k^2g(x,x_m)
h(x_m)\mathcal{P}_m+h(x_m)(\mathcal{P}_m,
\nb)\nb g(x,x_m)\}|_{x=x_j}
\ee
where $1\le j \le M,$ and
\be\label{eA6}\mathcal{P}_{0j}:=(\nb \times E_0)(x_j),\quad
\mathcal{P}_{j}=(\nb \times E)(x_j), \quad 1\le j\le M.
\ee
Equation \eqref{eA5} is a linear
algebraic system for finding $\mathcal{P}_m$.
In the above derivation we have used the known formula
$$\nabla\times[A,B]=(B, \nabla)A-(A,\nabla)B+A (\nabla,B)- B(\nabla,A),$$
which yields
$$\nabla\times [\nabla g,Q_m]=(Q_m,\nabla)\nabla g-Q_m\nabla^2 g,$$
where
$$(Q_m,\nabla)\nabla g=\sum_{i=1}^3 e_i\sum_{i'=1}^3
Q_{mi'}\frac{\partial^2 g}{\partial x_{i}\partial x_{i'}},\quad
Q_{mi}=(Q_m,e_i),$$
and $e_i$, $i=1,2,3,$ is the orthonormal Cartesian basis of $\R^3$.

\end{document}